\documentclass[preprint,aps,floatfix,nofootinbib,groupedaddress,superscriptaddress]{revtex4}
\usepackage[colorlinks,linkcolor=blue,anchorcolor=blue,citecolor=blue]{hyperref}
\usepackage{hyperref}
\usepackage{float}
\usepackage{amsfonts}
\usepackage{amssymb}
\usepackage{savesym}
\usepackage{amsmath}
\savesymbol{iint}
\usepackage{txfonts}
\restoresymbol{TXF}{iint}
\usepackage{ctable}

\usepackage{epsfig}
\usepackage{textcomp}
\usepackage[page]{appendix}
\usepackage{bm}
\usepackage{subfigure}
\usepackage[T1]{fontenc}
\usepackage[utf8]{inputenc}
\usepackage{babel}
\usepackage{booktabs}
\usepackage{threeparttable}

\begin{document}

\title{Realization of Heisenberg models of spin systems with polar molecules in pendular states}

\author{Wenjing Yue}\affiliation{State Key Laboratory of Precision Spectroscopy, East China Normal University, Shanghai, China}
\author{Qi Wei \footnote{ Corresponding email: qwei@admin.ecnu.edu.cn}} \affiliation{State Key Laboratory of Precision Spectroscopy, East China Normal University, Shanghai, China}
\author{Sabre Kais}\affiliation{Department of Chemistry, Purdue University, West Lafayette, Indiana 47907, USA}
\author{Bretislav Friedrich}\affiliation{Fritz-Haber-Institut der Max-Planck-Gesellschaft, Faradayweg 4-6, D-14195 Berlin, Germany}
\author{Dudley Herschbach}\affiliation{Department of Chemistry and Chemical Biology, Harvard University, Cambridge, Massachusetts, USA}

\begin{abstract}
We show that ultra-cold polar diatomic or linear molecules, oriented in an external electric field and mutually coupled by dipole-dipole interactions, can be used to realize the exact Heisenberg XYZ, XXZ and XY models without invoking any approximation. The two lowest lying excited pendular states coupled by microwave or radio-frequency fields are used to encode the pseudo-spin. We map out the general features of the models by evaluating the models’ constants as functions of the molecular dipole moment, rotational constant, strength and direction of the external field as well as the distance between molecules. We calculate the phase diagram for a linear chain of polar molecules based on the Heisenberg models and discuss their drawbacks, advantages, and potential applications.
\end{abstract}

\maketitle

\section{Introduction}

Introduced by Heisenberg in 1928, the Heisenberg statistical model of spin systems has been widely used to study phase transitions and critical phenomena in magnetic systems and strongly correlated electron systems\textsuperscript{\cite{1-Roger G.Bowers1969,2-J.F.Cooke1970,3-Freeman J.Dyson1976,4-Guang-Shan Tian1997,5-Henk W.J2002,6-R.G.Brown2006}}. Recently developed powerful tools developed to unravel the physics of strongly correlated multi-body quantum systems provide new platforms for understanding quantum magnetism\textsuperscript{\cite{7-Fernanda2013}}. It has been proposed to implement the Heisenberg model in other systems as well. For example, Pinheiro {\it et al.} demonstrated that in the Mott region, a boson atom in the first excitation band of a two-dimensional optical lattice can realize the spin-1/2 quantum Heisenberg model\textsuperscript{\cite{7-Fernanda2013}}. Bermudez {\it et al.} introduced a theoretical scheme to simulate the XYZ model using trapped ions\textsuperscript{\cite{8-A. Bermudez2017}}.

The molecular axis of polar molecules that are subject to an external electric field oscillates within a certain angular range about the field direction, forming pendular states\textsuperscript{\cite{9-B. Friedrich1991}}. These pendular states have specific orientations that give rise to constant projections of the dipole moment along the external field, resulting in long-range anisotropic interactions via the electric dipole-dipole coupling. In a field gradient, pendular molecule can be individually addressed due to its field-dependent eigenenergy (and orientation). Moreover, the internal structure of polar molecules is much richer than that of atoms or spins, allowing much richer physics. Given these unique properties, arrays of polar molecules are considered to be promising platforms for quantum computing and quantum information processing\textsuperscript{\cite{10-Book2009,11-D. DeMille2002,12-Philippe Pellegrini2011,13-Jing Zhu2013,14-Zuo-Yuan Zhang2017,15-Zuo-Yuan Zhang2020,16-Wei12011,17-Wei22011,32-Micheli,33-Charron,34-kuz,35-ni,36-lics,37-YelinDeMille,38-Wei2010,39-Wei2016,40-kang-Kuen Ni2018}}, which is not unlike spins.

Inspired by the similarity between spins and polar molecules, the simulation of the spin models with polar molecules has attracted broad interests over the past decade\textsuperscript{\cite{18-Muller2010,19-Alexey2011,20-Bo Yan2013,21-N.Y.Yao2018,22-Haiyuan Zou2017,22+-Kaden2013,M. L. Wall,A. V. Gorshkov}}. M{\"u}ller described in his thesis the details of how to realize the spin-1/2 XXZ model as well as t-J model with ultra-cold polar molecules trapped in an optical lattice\textsuperscript{\cite{18-Muller2010}}. Gorshkov {\it et al.} demonstrated that the dipole interactions of ultra-cold alkali metal dimers in optical lattices can be used to implement the t-J model, providing insights into strong correlation phenomena in condensed systems\textsuperscript{\cite{19-Alexey2011}}. Yan {\it et al.} experimentally observed dipolar spin-exchange interactions with lattice-confined polar molecules, which laid a foundation for further study of multi-body dynamics in spin lattices\textsuperscript{\cite{20-Bo Yan2013}}. Yao {\it et al.} obtained the dipole Heisenberg model by using polar molecules and found the existence of quantum spin liquids on the triangular and Kagome lattice\textsuperscript{\cite{21-N.Y.Yao2018}}. Zou {\it et al.} implemented the quantum spin model based on the polar molecule KRb in an optical lattice and discovered the quantum spin liquid on the square lattice\textsuperscript{\cite{22-Haiyuan Zou2017}}.

However, in almost all the previous works about implementation of the spin-1/2 Heisenberg model with polar molecules, the ground and first excited pendular states with $M=0$ were usually chosen as pseudo-spin states, representing spin up and spin down, respectively\textsuperscript{\cite{18-Muller2010,19-Alexey2011,20-Bo Yan2013,21-N.Y.Yao2018,22-Haiyuan Zou2017,22+-Kaden2013}}. In which case the Hamiltonian is not in the form of the Heisenberg model. Only after applying the rotating wave approximation can the Heisenberg model be recovered. Furthermore, it is not a general Heisenberg XYZ model, but its special case, the XXZ model.

Herein, by choosing the two lowest excited pendular states of a polar molecule to represent the pseudo-spin states, we show how to achieve spin-1/2 Heisenberg XYZ model as well as XXZ and XY models directly, without any approximation. We work out the properties of the models by evaluating all their constants as functions of three dimensionless variables.  The first one is $\mu\varepsilon/B$, the ratio of the Stark energy (magnitude of permanent dipole moment times electric field strength) to the rotational constant (proportional to inverse of the molecular moment of inertia); this variable governs the energy and intrinsic angular shape of the pendular states. The second one is $\Omega/B$, with $\Omega$ = $\mu^2/r^3$, the square of the permanent dipole moment divided by the cube of the separation distance; this variable governs the magnitude of the dipole-dipole coupling. The third variable is $\alpha$, the angle between the axis of the molecular array and the electric field. As a sample application of the Heisenberg model based on polar molecules, we construct the ground state phase diagram for a linear array of polar molecules. We also discuss advantages, drawbacks as well as potential applications of our model.

\section{Pendular states of polar molecules as pseudo-spins} \label{2}

\subsection{Pendular and pseudo-spin states}

\begin{figure}[htp]
 \vspace{0.5cm}
\setlength{\abovecaptionskip}{-0.2cm}
\setlength{\belowcaptionskip}{0cm}
\centering
\includegraphics[width=0.6\columnwidth]{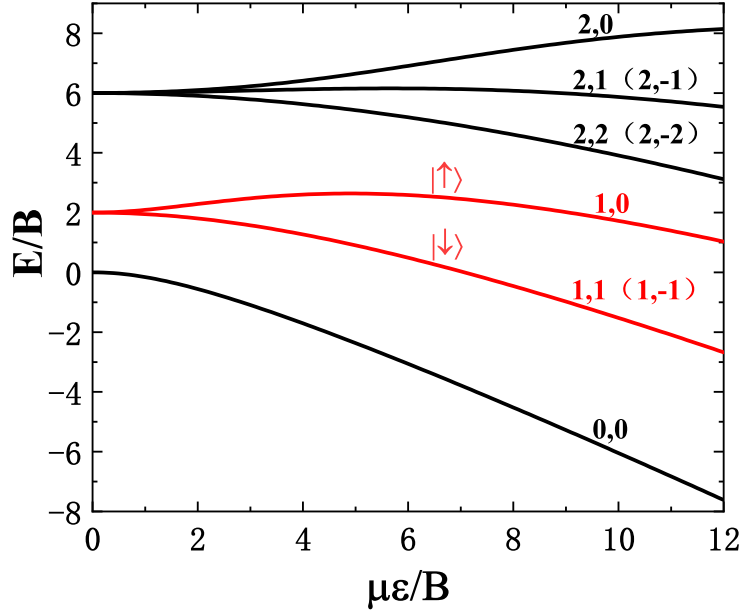}
\caption{Eigenenergies of a polar molecule in an external electric field, as functions of $\mu \varepsilon /B$, with $\mu $ the permanent dipole moment, $\varepsilon $ the field strength, {\it B} the rotational constant.  $\left|  \downarrow  \right\rangle $ correlates with the $J = 1$, ${M} = 1$ and $\left|  \uparrow  \right\rangle $ with the $J = 1$, ${M} = 0$ states. States used as the pseudo-spin states (red curves) are labeled $\left|  \downarrow  \right\rangle $ and $\left|  \uparrow  \right\rangle $ in the absence of an external field. }
\label{fig1}
\end{figure}

In an electrostatic field, the Hamiltonian of a trapped linear polar molecule is \textsuperscript{\cite{16-Wei12011}}
\begin{equation}
H = \frac{{{p^2}}}{{2m}} + {V_{trap}}({\bf r}) + B{{\bf J}^2} - \boldsymbol{\mu}  \cdot \boldsymbol{\varepsilon},
\end{equation}
where the molecule, with mass $m$, rotational constant $B$, and body-fixed dipole moment $\mu$, has translational kinetic energy $p^2/2m$, potential energy $V_{trap}$ within the trapping field and rotational energy $B\bf{J}^2$ as well as interaction energy $\boldsymbol{\mu}  \cdot \boldsymbol{\varepsilon}$ with the external field $\boldsymbol{\varepsilon}$. In the trapping well, at ultra-cold temperatures, the translational motion of the molecule is quite modest and very nearly harmonic; $p^2/2m + V_{trap}(r)$ is thus nearly constant and can be omitted from the Hamiltonian. There remains the rotational kinetic energy and Stark interaction,
\begin{equation}
H_s=B{\bf J}^2-\mu\varepsilon\cos\theta,
\end{equation}
where $\theta$ is the polar angle between the molecular axis (the molecule-fixed permanent electric dipole moment $\mu$) and the field direction. Under the action of a strong electrostatic field, the polar molecules are compelled to undergo pendular oscillations and result in the forming of pendular states, $\vert$$\tilde{J}M\rangle$. Here, $\tilde{J}$ wears a tilde to indicate it is no longer a good quantum number since the Stark interaction mixes the rotational states, whereas $M$ is still a good quantum number as long as azimuthal symmetry about $\varepsilon$ is maintained. Figure \ref{fig1} shows eigenenergies of a few lowest lying pendular states for a $^1\Sigma$ diatomic (or linear)
molecule, as functions of $\mu \varepsilon /B$.

We choose the two lowest excited states pendular states, $\vert$11$\rangle$ and $\vert$10$\rangle$, as the pseudo-spin states $\left|  \downarrow  \right\rangle$ and $\left|  \uparrow  \right\rangle $, respectively (see Figure \ref{fig1}). Then use an external  circularly polarized microwave or radio-frequency field to couple the two states, forming a $\left|  \downarrow  \right\rangle$ and $\left|  \uparrow  \right\rangle $ two-level system. The two pseudo-spin states  are linear superpositions of spherical harmonics $Y_j^1$and $Y_j^0$:
\begin{equation}
\vert\downarrow\rangle=\sum_{\substack{j}}a_jY_j^1(\theta,\phi)\qquad\vert\uparrow\rangle=\sum_{\substack{j}}b_jY_j^0(\theta,\phi).
\end{equation}

\begin{figure}[htp]
 \vspace{0.5cm}
\setlength{\abovecaptionskip}{-0.2cm}
\setlength{\belowcaptionskip}{0cm}
\centering
\includegraphics[width=0.7\columnwidth]{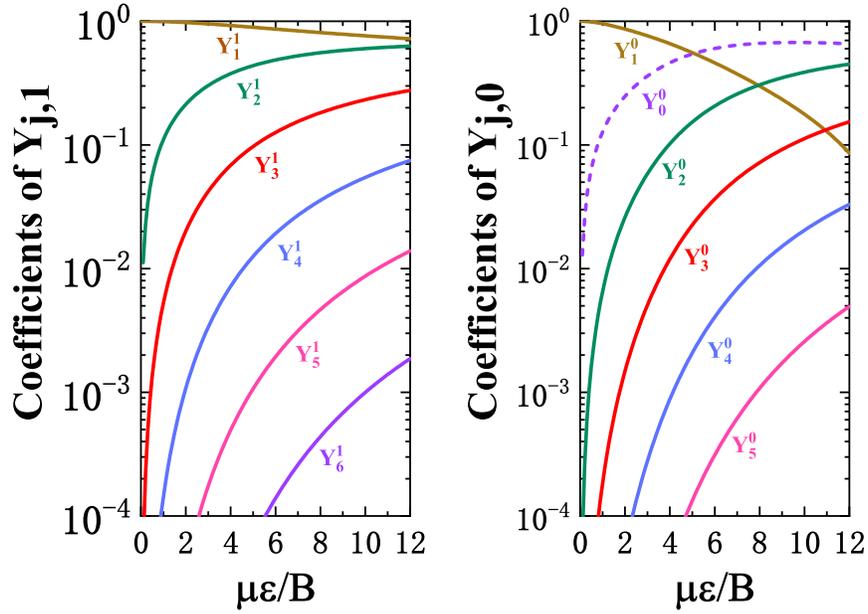}
\caption{Coefficients of spherical harmonics for pendular states $\left|\downarrow\right\rangle$ (left panel) and $\left|\uparrow\right\rangle$ (right panel), see Eq.(3). Dashed curve for $\left|\uparrow\right\rangle $ indicates the coefficients of $Y_0^0$ is negative. }
\label{fig2}
\end{figure}

\begin{figure}[htp]
 \vspace{0.5cm}
\setlength{\abovecaptionskip}{-0.2cm}
\setlength{\belowcaptionskip}{0cm}
\centering
\includegraphics[width=1\columnwidth]{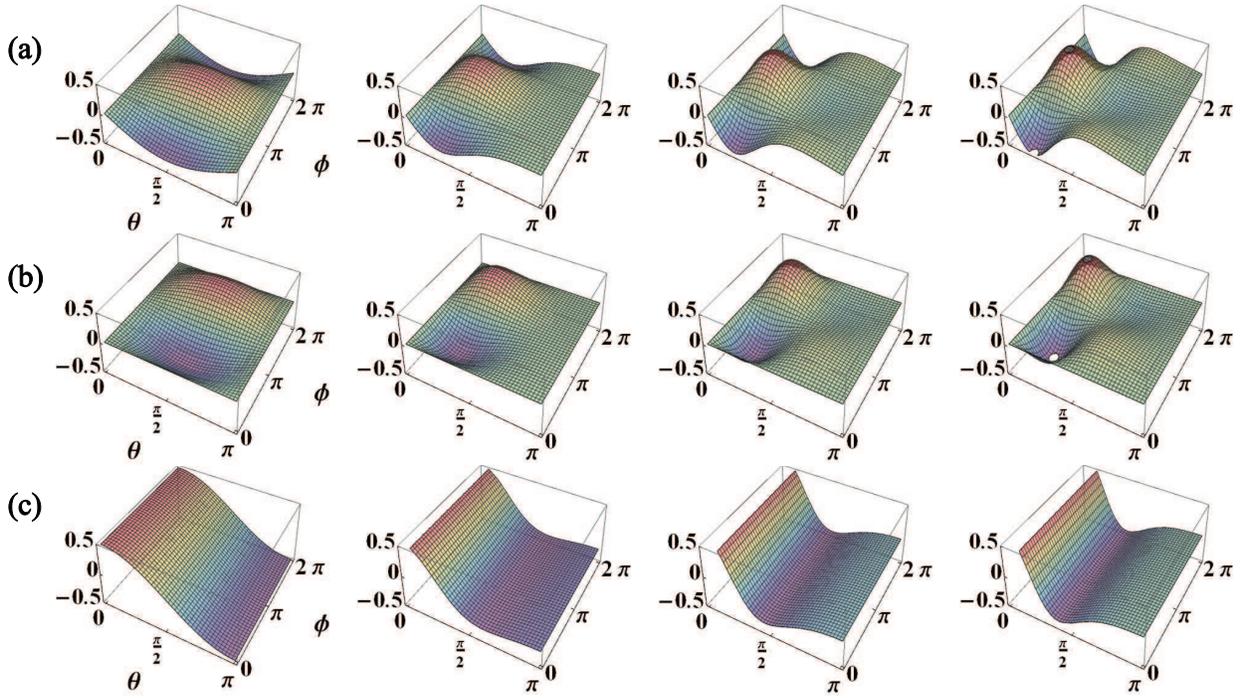}
\caption{Wave functions of the $\left|\downarrow\right\rangle$ and $\left|\uparrow\right\rangle$ pendular states for values of $\mu \varepsilon/B = 0,4,8,12$ (from left to right), respectively. Panels (a) and (b) represent the real and imaginary parts of state $\left|\downarrow\right\rangle$ respectively. Panel (c) represents state $\left|\uparrow\right\rangle$ (has no imaginary part).}
\label{fig3}
\end{figure}

Figure \ref{fig2}  plots the coefficients as functions of $\mu\varepsilon /B$. For $\mu\varepsilon /B$ = 0, both $\left| \downarrow \right\rangle$ and $\left| \uparrow \right\rangle$ are purely rotational states with single component of spherical harmonics of $Y_1^1$ and $Y_1^0$, respectively. As  $\mu\varepsilon /B$ increases, more and more components of spherical harmonics with the same $M$ but different $J$ get involved and the initially dominant components decrease accordingly. For $\left| \uparrow \right\rangle$, the dominant component $Y_1^0$ (shown in brown) decreases so quickly that it is replaced by $Y_0^0$ as the leading term for $\mu\varepsilon /B
>4.5$.  For $\left| \downarrow \right\rangle$, the initially dominant component $Y_1^1$ (show in brown) decreases a little slower but is eventually replaced by $Y_1^1$ when $\mu\varepsilon /B$ becomes large enough. Figure \ref{fig3} displays wave functions of $\left| \downarrow \right\rangle$ and $\left| \uparrow \right\rangle$ for different magnitudes of the electric field.  For $\left| \uparrow \right\rangle$, since $M=0$, the dipole is rotating with its ${\mathbf J}$-vector perpendicular to the field direction. Without the external field, the dipole orientation is symmetric in the hemispheres toward ($\theta = 0$) or opposite  ($\theta =\pi$) to the field direction. With increasing external field, the pinwheeling dipole favors the opposite hemisphere because its motion is slowed down there. However, when the external field becomes large enough, pinwheeling is inhibited and converted into pendular libration about the field direction, and the dipole orientation favors the toward hemisphere. For $\left| \downarrow \right\rangle$, since $M=1$, without the external field, the angular momentum is along the field direction, and thus the dipole orientation is localized at about $\theta = \pi/2$ and also symmetric in both hemispheres toward or opposite to the field direction. As the external field increases, the dipole rotates like a conical pendulum and its orientation favors more and more the toward hemisphere.

\subsection{Hamiltonian of psedo-spins with electric dipole-dipole interaction}\label{3}

Adding a second trapped polar molecule, identical to the first one but a distance $r_{12}$ apart, introduces the dipole-dipole interaction term\textsuperscript{\cite{16-Wei12011}}
\begin{equation}
{V_{d - d}} = \frac{{{\bm{\mu _1}} \cdot {\bm{\mu _2}} - 3\left( {{\bm{\mu _1}} \cdot \bm{n}} \right)\left( {{\bm{\mu _2}} \cdot \bm{n}} \right)}}{{{{\left| {{\bm{r_1}} - {\bm{r_2}}} \right|}^3}}}.
\end{equation}
Here {\bf n} is a unit vector along ${\bf r}_{12}$. In the presence of an external field, $V_{d-d}$ can be expressed in terms of the polar and azimuthal angles:
\begin{equation}
\begin{aligned}
V_{d-d}=&\Omega\left[\cos\theta_1\cos\theta_2+\sin\theta_1\cos\varphi_1\sin\theta_2\cos\varphi_2+\sin\theta_1\sin\varphi_1\sin\theta_2\sin\varphi_2\right.\\&\left.-3\left(\sin\theta_1\cos\varphi_1\sin\alpha+\cos\theta_1\cos\alpha\right)\left(\cos\theta_2\cos\alpha+\sin\theta_2\cos\varphi_2\sin\alpha\right)\right],
\end{aligned}
\end{equation}
where $\Omega  = {\mu ^2}/r_{12}^3$, $\alpha$ is the angle between the ${\bf r}_{12}$ vector and the field direction, $\theta_1$ and $\theta_2$ are the polar angles between the dipoles (${\bm \mu}_1$ and ${\bm\mu}_2$) and the field direction, $\varphi_1$ and $\varphi_2$ are the corresponding azimuths.

Now the total Hamiltonian is ${H_{total}}$ = ${H_{s1}}$ + ${H_{s2}}$ + ${V_{d - d}}$. When set up in the basis set of the direct product of pseudo-spin states $\{|\downarrow\downarrow\rangle,|\downarrow\uparrow\rangle,|\uparrow\downarrow\rangle,|\uparrow\uparrow\rangle\}$, it takes the form
\begin{equation}
{H_{s1}} + {H_{s2}} = \left( {\begin{array}{*{20}{c}}{2{E_0}}&0&0&0\\0&{{E_0} + {E_1}}&0&0\\0&0&{{E_1} + {E_0}}&0\\0&0&0&{2{E_1}}\end{array}} \right),
\end{equation}
\begin{equation}
{V_{d - d}} = \Omega \left( {\begin{array}{*{20}{c}}{{P_\alpha }C_0^2}&0&0&{{Q_\alpha }C_X^2}\\0&{{P_\alpha }{C_0}{C_1}}&{ - {P_\alpha }C_X^2}&0\\0&{ - {P_\alpha }C_X^2}&{{P_\alpha }{C_1}{C_0}}&0\\
{{Q_\alpha }C_X^2}&0&0&{{P_\alpha }C_1^2}\end{array}} \right),
\end{equation}
where $E_0$ and $E_1$ are eigeneneries of the pendular pseudo spin states $\vert$$\downarrow$$\rangle$ and $\vert$$\uparrow$$\rangle$, respectively (see Figure \ref{fig1}). $P_\alpha$ and $Q_\alpha$ are simple functions of $\alpha$: ${P_\alpha } = 1 - 3{\cos ^2}\alpha$, ${Q_\alpha } =  - 3{\sin ^2}\alpha $. In $V_{d-d}$, the basis states are linked by matrix elements containing $C_0 $ and $C_1$, the field-induced dipole moments orientation cosines, and $C_X$, the transition dipole moments between the pseudo-spin states $\vert\downarrow\rangle$ and $\vert\uparrow\rangle$. These are given by
\begin{equation}
C_0=\langle\downarrow\vert\cos\theta\vert\downarrow\rangle\qquad C_1=\langle\uparrow\vert\cos\theta\vert\uparrow\rangle\qquad C_X=\langle\downarrow\vert\sin\theta\cos\varphi\vert\uparrow\rangle.
\end{equation}

\begin{figure}[htp]
 \vspace{0.5cm}
\setlength{\abovecaptionskip}{-0.2cm}
\setlength{\belowcaptionskip}{0cm}
\centering
\includegraphics[width=0.6\columnwidth]{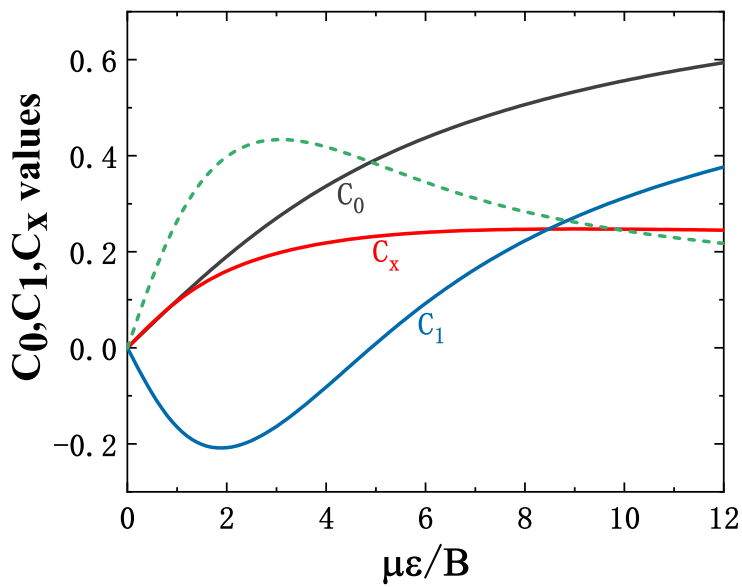}
\caption{Matrix elements of $C_0$, $C_1$ and $C_X$ as functions of  $\mu\varepsilon/B$. The dotted green line is ${C_0} - {C_1}$.}
\label{fig4}
\end{figure}

In contrast to a real spin state which has a constant dipole moment, here, the values of $C_0 $, $C_1$ and $C_X$  are functions of external electric fields, which are displayed in Figure \ref{fig4}. When $\mu \varepsilon /B$ increases, $C_0$ becomes increasingly positive, whereas $C_1$ is increasingly negative upto about  $\mu\varepsilon/B=2$, then climbs to zero at about  $\mu\varepsilon/B=4.9$ and thereafter is increasingly positive. The fact that $C_X = 0$ at  $\mu\varepsilon/B$ = 0 means that without the external electric field, the transition between $\left|\downarrow\right\rangle$ and $\left|\uparrow\right\rangle$ is not allowed as a one-photon electric dipole transition. Fortunately, increasing the external field introduces sufficient mixing of other spherical harmonics, particularly admixing of $Y_0^0$ and $Y_2^0$ into $\left|\uparrow\right\rangle$ and admixing of $Y_2^1$ into $\left|\downarrow\right\rangle$ (see Figure \ref{fig2}), such that $C_X$ increases sharply from zero to a considerable value, enabling the $\left|\downarrow \right\rangle\leftrightarrow\left|\uparrow\right\rangle$ transition to occur as a one-photon transition.

\section{Realization of the Heisenberg model of spin systems with polar molecules.}

\subsection{General Heisenberg XYZ model based on pseudo-spins.}

The total Hamiltonian of the two-dipoles molecular system can be mapped onto a two-qubit spin-1/2 general Heisenberg XYZ model:
\begin{equation}
{H_{XYZ}} = {J_x}\sigma _1^x\sigma _2^x + {J_y}\sigma _1^y\sigma _2^y + {J_z}\sigma _1^z\sigma _2^z - \gamma\left( {\sigma _1^z + \sigma _2^z} \right),
\label{eq9}
\end{equation}
here ${\sigma _x}$, ${\sigma _y}$ and ${\sigma _z}$ are Pauli operators; $J_x, J_y, J_z$ and $\gamma$ are coupling constants given by
\begin{equation}
\begin{split}
{J_x}&=\Omega \left( {3{{\cos }^2}\alpha  - 2} \right)C_X^2, \\
{J_y}&=\Omega C_X^2, \\
{J_z} &= \frac{{\Omega \left( {1 - 3{{\cos }^2}\alpha } \right){{\left( {{C_0} - {C_1}} \right)}^2}}}{4},\\
 \gamma &  = \frac{{2\left( {{E_1} - {E_0}} \right) + \Omega \left( {3{{\cos }^2}\alpha  - 1} \right)\left( {C_0^2 - C_1^2} \right)}}{4}.
\end{split}
\label{eq10}
\end{equation}

\begin{figure}[htbp]
\vspace{0.5cm}
\setlength{\abovecaptionskip}{-0.2cm}
\setlength{\belowcaptionskip}{0cm}
\centering
\includegraphics[width=0.6\columnwidth]{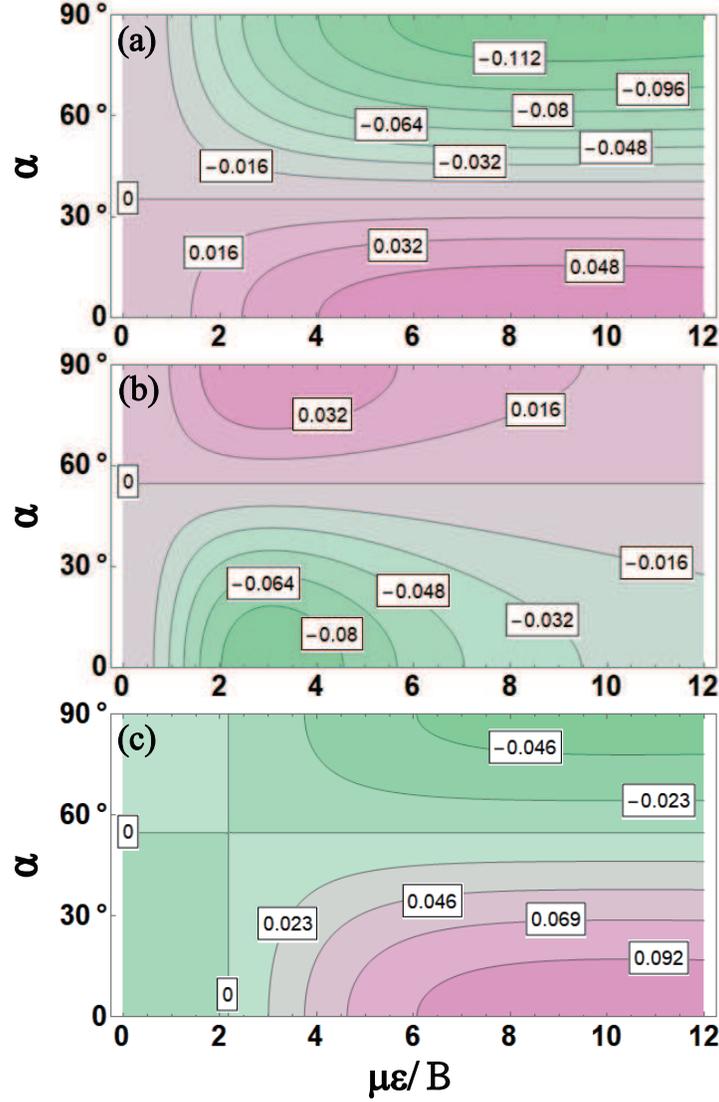}
\caption{Contour plots of $J_x$$/\Omega$ (panel a), $J_z$$/\Omega$ (panel b) and second part of $\gamma$$/\Omega$ (panel c) as functions of reduced variable $\mu \varepsilon /B$ and angle $\alpha$. $J_y$ is $\alpha$-independent and the same as $J_x$ when $\alpha=0$.}
\label{fig5}
\end{figure}

Equations (\ref{eq9}-\ref{eq10}) demonstrate how to realize the spin-1/2 anisotropic Heisenberg model with polar molecules in pendular states. The model constants ($J_x, J_y, J_z$ and $\gamma$) are functions of  $\mu \varepsilon /B$, $\Omega$ and $\alpha$, which means the model can be adjusted by modifying these parameters. For all the constants, the relations to $\Omega$ are simply linear, whereas the relations to $C$'s are quadratic. $J_y$ is $\alpha$ independent and equal to $J_x$ with $\alpha=0$. The term $\gamma$ consists of two parts. One part is related to the energy gap $\Delta E=E_1-E_0$ which is shown in Figure \ref{fig1}. The other part is proportional to $\Omega$ which is similar to $J$'s. The contour plots in Figure \ref{fig5} illustrate how  $J_x/\Omega$, $J_z/\Omega$ and the second part of $\gamma/\Omega$ change with $\mu\varepsilon/B$ and $\alpha$. When $\mu\varepsilon/B$ increases from 0 to 12, the magnitude of the coupling coefficients ($J_x/\Omega$, $J_y/\Omega$ and $J_z/\Omega$) changes in the order of 0 to $10^{-1}$. Similar results are obtained for the second part of $\gamma$$/\Omega$ . Maximum or minimum values of $J_x$ and $J_y$ appear at large $\mu\varepsilon/B$, whereas for $J_z$ they appear around $\mu\varepsilon/B=3$.

\begin{figure}[htbp]
\vspace{0.5cm}
\setlength{\abovecaptionskip}{-0.2cm}
\setlength{\belowcaptionskip}{0cm}
\centering
\includegraphics[width=0.6\columnwidth]{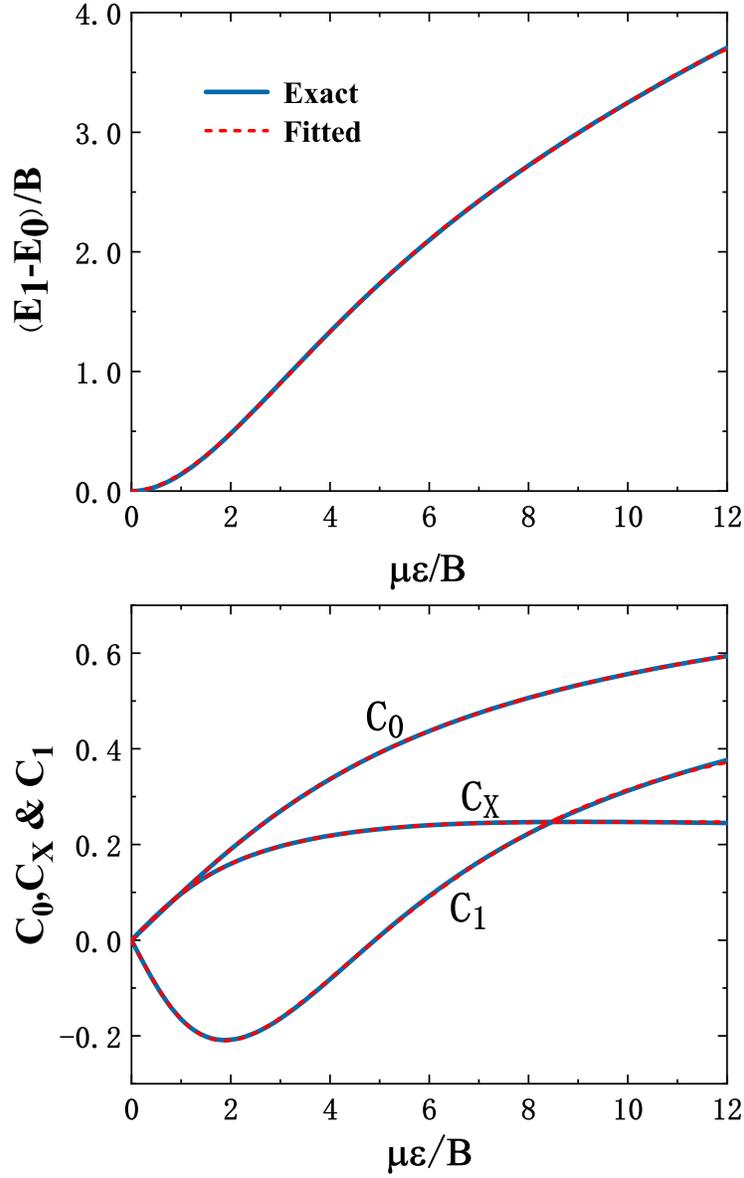}
\caption{Comparison of exact results (blue curves) with fitted approximation functions (dashed red curves) cf. Eqs. (11) and (12) : energy difference, ${{\left( {{E_1} - {E_0}} \right)} \mathord{\left/{\vphantom {{\left( {{E_1} - {E_0}} \right)} B}} \right. \kern-\nulldelimiterspace} B}$, the field-induced dipole moments, ${C_0}$ and ${C_1}$ and the transition dipole moments, ${C_X}$.}
\label{fig6}
\end{figure}

For given $\Omega $ and $\alpha $, the coupling constants ${J_x}$, ${J_y}$, ${J_z}$ and $\gamma$ depend only on  $x=\mu \varepsilon /B$, which enters those constants through  ${C_0}$, ${C_1}$, ${C_X}$ and $\Delta E$. To provide a convenient means to evaluate Equation (\ref{eq10}), we fitted our numerical results to obtain accurate approximation formulas,
\begin{align}
\label{eq11} & {{\left( {{E_1} - {E_0}} \right)} \mathord{\left/{\vphantom {{\left( {{E_1} - {E_0}} \right)} B}} \right.\kern-\nulldelimiterspace} B} = {A_1}x + {A_2}{x^2} + {A_3}{x^3} + {A_4}{x^4} + {A_5}{x^5},\\
\label{eq12} &C(x) = {A_0} + \frac{{{A_1}}}{{1 + \exp [(x - {x_1})/{k_1}]}} + \frac{{{A_2}}}{{1 + \exp [ - (x - {x_2})/{k_2}]}}.
\end{align}
These functions are plotted in Figure {\ref{fig6}}. The fitted parameters are given in Tables (\ref{Table1}-\ref{Table2}).

\begin{table}[!ht]
\renewcommand\tabcolsep{61pt}
\caption{Values of the parameters for Eq.{\ref{eq11}}.}
\label{Table1}
\centering
\begin{threeparttable}
\begin{tabular}{*4{c}}
  \hline
  \hline
Parameters & Values  \\
  \hline
 ${A_1}$ & 0.00794  \\
 ${A_2}$ & 0.16531  \\
 ${A_3}$ & -0.02838 \\
 ${A_4}$ & 0.00206  \\
 ${A_5}$ & $ - 5.55762 \times {10^{ - 5}}$ \\
  \hline
  \hline
\end{tabular}
\begin{tablenotes}
\footnotesize
\item ${R^2} = 0.9999$.
\end{tablenotes}
\end{threeparttable}
\end{table}

\begin{table}[!ht]
\renewcommand\tabcolsep{18pt}
\caption{Values of the parameters for Eq.{\ref{eq12}}.}
\label{Table2}
\centering
\begin{threeparttable}
\begin{tabular}{*4{c}}
  \hline
  \hline
Parameters & Values for ${C_0}$ & Values for ${C_X}$  &Values for ${C_1}$\\
  \hline
 ${A_0}$& -0.24612 & 0.21844 & -0.91801 \\
${A_1}$& -0.56893 & -0.53637 & 0.9 \\
${A_2}$& 0.95967 & 0.02855 & 1.36773 \\
${x_1}$& -0.09066 & -04403 & 0.09317 \\
${x_2}$& -1.25815 & 4.28747 & 2.52364 \\
${k_1}$& 2.17868 & 1.18595 & 0.80729 \\
${k_2}$& 6.7313 & 0.94214 & 3.38213 \\
  \hline
  \hline
\end{tabular}
\begin{tablenotes}
\footnotesize
\item ${R^2} = 1$ for ${C_0}$ , ${R^2} = 0.9999$ for ${C_X}$ and ${C_1}$.
\end{tablenotes}
\end{threeparttable}
\end{table}

The Heisenberg model given by Equations (\ref{eq9}-\ref{eq10}) is a general XYZ model. But we can get two different special cases of the Heisenberg model by changing the direction of the external field. One is the XXZ model which obtains by taking $\alpha  = {0^ \circ }$. In that case, we have ${J_x} = {J_y} \ne {J_z}$ and ${J_x} = {J_y} \ne 0$, ${J_z} \ne 0$. The other one is the XY model which obtains for $\alpha  = {54.7^ \circ }$, known as the magic angle. In that case, we have  ${J_x} \ne 0$, ${J_y} \ne 0$ and ${J_z} = 0$.

\subsection{The Heisenberg XXZ model and quantum phase diagram of polar molecules.}

\begin{figure}[htbp]
\vspace{-0.5cm}
\setlength{\abovecaptionskip}{-0.2cm}
\setlength{\belowcaptionskip}{0cm}
\centering
\includegraphics[width=0.6\columnwidth]{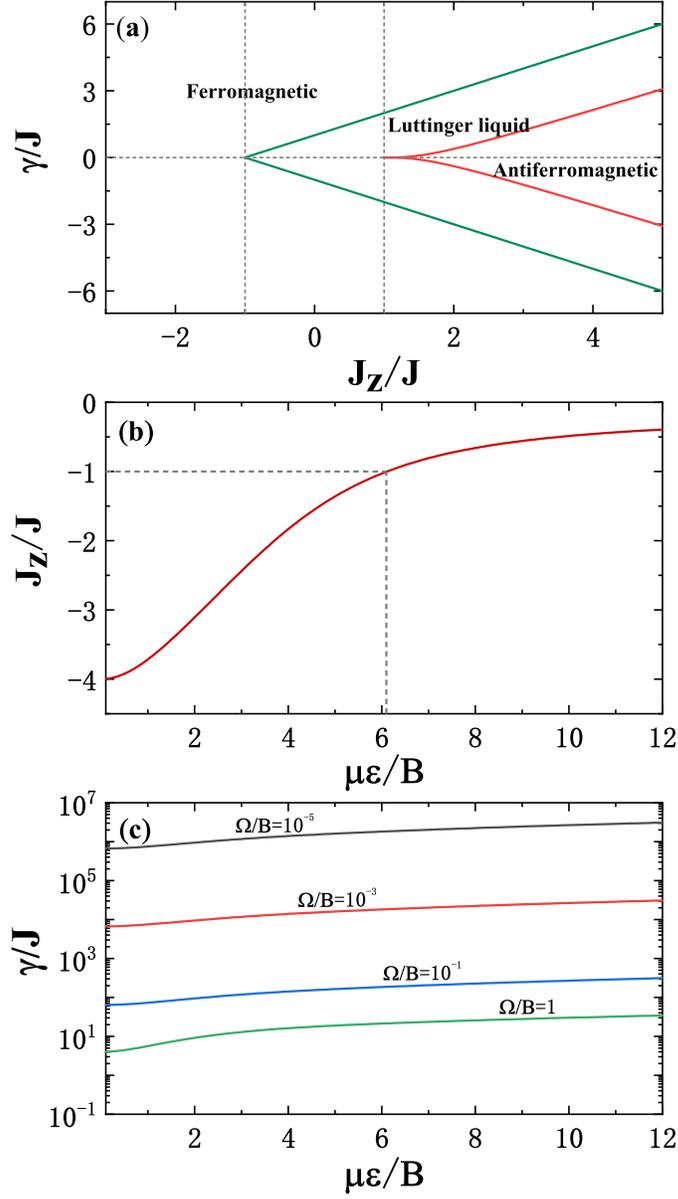}
\caption{(a) Quantum phase diagram of the XXZ model associated with $J_z/J$ and $\gamma/J$ for a linear spin chain. (b) The ratio of the coupling constants of the XXZ model, $J_z/J$, as a function of  $\mu\varepsilon/B$ in dipole system of polar molecules. (c) The ratio of the coupling constants of the XXZ model, $\gamma/J$, as a function of reduced variables, $\mu\varepsilon/B$ and $\Omega/B$, in dipole system of polar molecules.}
\label{fig7}
\end{figure}

In order to demonstrate the application of the Heisenberg model based on pendular  polar molecules, we take the XXZ model ($\alpha  = {0^ \circ }$) as an example. If only pairwise interaction is considered, for a system with $N$ polar molecules trapped in a linear array with external electric field along the array, the Hamiltonian has the form of the XXZ model,
\begin{equation}
{H_{XXZ}} = \sum\limits_{i = 1}^{N-1} {\left[ {J\left( {\sigma _i^x\sigma _{i+1}^x + \sigma _i^y\sigma _{i+1}^y} \right)} + {J_z}\sigma _i^z\sigma _{i+1}^z\right]} - \gamma\sum\limits_{i = 1}^{N}{\sigma _i^z}
\end{equation}
with couplings given by
\begin{equation}
\begin{split}
 J&=\Omega C_X^2, \\
 {J_z} &=- \frac{{\Omega {{\left( {{C_0} - {C_1}} \right)}^2}}}{2},\\
 \gamma &  = \frac{{\left( {{E_1} - {E_0}} \right) + \Omega \left( {C_0^2 - C_1^2} \right)}}{2}.
\end{split}
\label{eq16}
\end{equation}

\begin{figure}[htbp]
 \vspace{0.5cm}
\setlength{\abovecaptionskip}{-0.2cm}
\setlength{\belowcaptionskip}{0cm}
\centering
\includegraphics[width=0.6\columnwidth]{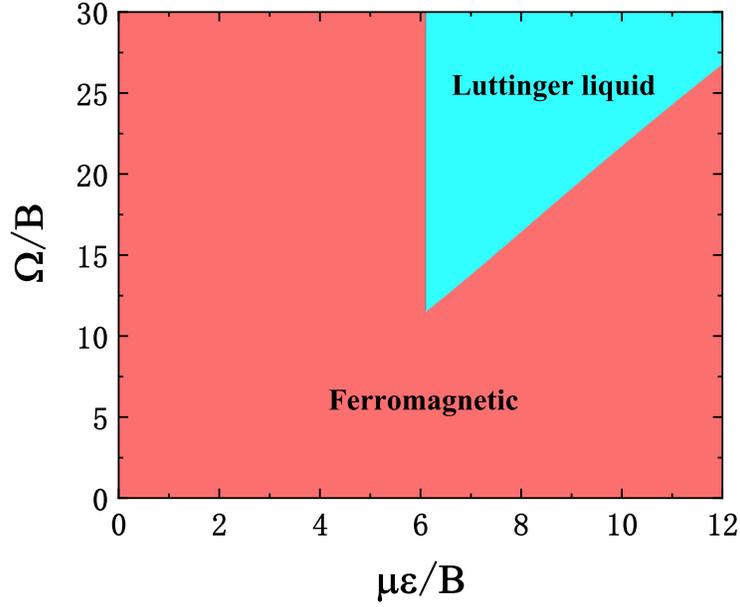}
\caption{Ground state phase diagram of the XXZ model associated with $\Omega /B$ and $\mu \varepsilon /B$ for polar molecules in a linear array.}
\label{fig8}
\end{figure}

Figure \ref{fig7}(a) displays the ground state phase diagram of a spin-1/2 XXZ chain with nearest-neighbor interaction\textsuperscript{\cite{23-Christian2010,24-Mykhailo2019,25-D. C. Cabra1998}}. The abscissa is the scaled anisotropy parameter $J_z/J$, and the ordinate is the scaled magnetic field $\gamma/J$. There are two gapped phases: one is the ferromagnetic phase for $J_z/J < -1$; the other is the antiferromagnetic phase for $J_z/J > 1$. In between is the Luttinger liquid phase\textsuperscript{\cite{26-F.D.M. Haldane1980}}. According to Equation \ref{eq16}, for polar molecules, $J_z/J$ only depends on $\mu\varepsilon/B$, so in Figure \ref{fig7}(b) we show how $J_z/J$ changes when $\mu\varepsilon/B$ increases from 0 to 12. The critical value of $J_z/J = -1$ appears at $\mu\varepsilon/B = 6.1 $ ($\varepsilon$ =13.5 kV/cm for the SrO molecule). In order to obtain phase information about the polar molecule system, we still need to know the range of values of $\gamma/J$. According to Equation \ref{eq16}, $\gamma/J$ depends on both $\mu\varepsilon/B$ and $\Omega$. So in Figure \ref{fig7}(c) we plot of $\gamma/J$ as a function of $\mu\varepsilon/B$ for different $\Omega/B$. Finally we obtain a ground state phase diagram associated with $\mu\varepsilon/B$ and $\Omega/B$ for a linear array of polar molecules, which is shown in Figure \ref{fig8}.

\section{Discussion and Prospects}\label{4}

In this paper, our chief aim was to demonstrate that the Heisenberg model of spin systems can be realized with ultra-cold diatomic or linear $^1\sum$ molecules, oriented in an external electrostatic field and coupled by the electric dipole-dipole interaction. This requires use of pendular states comprised of superpositions of spherical harmonics. Here the two lowest lying excited states coupled by microwave or radio-frequency fields are used to mimic the two-level spin system. This provides a new physical platform for the study of the Heisenberg model. Since the dipole is encoded in the rotational states of the molecules, the field-induced electric dipole-dipole interactions between the molecules reproduce magnetic dipole-dipole interactions between spins. In order to map out the general features of the model, we have considered a wide range of parameters defined by sets of unitless reduced variables, involving the dipole moments, rotational constant, dipole-dipole coupling, electric field strength and direction.

The external field plays an essential role. In order to induce extensive hybridization of rotational states, the field strength needs to be sufficiently high. This has a dual purpose. Firstly, to make the molecules undergo pendular oscillations about the field direction; otherwise rotational tumbling would average out the molecule's dipole moment in the laboratory frame. Secondly, to make the transition dipole moment $C_X$ deviate from zero such that a one-photon transition $\left|\downarrow\right\rangle\leftrightarrow\left|\uparrow\right\rangle$ would be fully allowed.

Using optical lattices to trap the molecules limits the distance between adjacent molecules to a few hundreds nanometers, so that the dipole-dipole coupling is weak ($\Omega/B$ typically of order $10^{-6}$ to $10^{-4}$) compared with the energy gap $\Delta E$ and thus the model parameter $\gamma/J$ becomes very large ($> 10^4$). For that weak coupling realm, the ground state of the Heisenberg model obtained with polar molecules is always in the ferromagnetic phase (see Figure \ref{fig8}). In order to enhance the dipole-dipole coupling, the molecular distance $r$ has to be shortened. For the SrO molecule as an example, the molecular distance of less than 10 nm is required for $\gamma/J \sim 1$. Such distance is much shorter than what can be achieved in typical optical lattices, but might be obtained with arrays of nanoscale plasmon-enhanced electro-optical traps\textsuperscript{\cite{27-Brian Murphy2009,28-D.E.Chang2009}} or molecular Wigner crystals\textsuperscript{\cite{29-P. Rabl2007,30-H. P. Buchler2007}}. That will be a promising approach to extend the experimental scope of the model.

Polar molecules also offer significant advantages for achieving the Heisenberg spin model, due to their high controllability and the presence of strong and long range interactions. Stark energy is quite large, so for instance the energy gaps between pseudo-spin states $\left|  \downarrow  \right\rangle$ and $\left|  \uparrow  \right\rangle$ are typically in the range of microwave frequencies, as opposed to the radiofrequencies separating of real spin states. This enables a faster optically controlled transition between the two energy levels of polar molecules. Electric dipole-dipole interaction is also much stronger than that of spins, resulting in a larger frequency shift, which is essential for building quantum logic gates.

The influence of the $\left| {1-1} \right\rangle $ pendular state  which is degenerate with $\left| {1 1} \right\rangle $ was also numerically analyzed. If we use $\left| {1-1} \right\rangle $ instead of $\left| {1 1} \right\rangle $ as the pseudo spin states $\left|  \downarrow  \right\rangle$, we will obtain the same field-induced dipole moments ${C_0}$, ${C_1}$ and the transition dipole moment ${C_X}$. That indicates that the pseudo-spin state  $\left|  \downarrow  \right\rangle$ can also be $\left| {1 -1} \right\rangle$, since the Hamiltonian matrix is the same as that for $\left| {1 1} \right\rangle$. But if the pseudo spin state  $\left|  \downarrow  \right\rangle$ is a superposition of $\left| {1 - 1} \right\rangle $ and $\left| {1 1} \right\rangle $, ${C_0}$ and ${C_1}$ will remain unchanged, whereas ${C_X}$ will be different. In this case, the Hamiltonian is still in the form of Heisenberg model, but the Hamiltonian matrix elements related to ${C_X}$ take on different values. Including the $\left| {1-1} \right\rangle $ state would increase the flexibility and complexity of the model, and will not be elaborated upon here. However, this problem can be avoided by introducing a tilt angle $\beta$ between the polarization vector of the optical trapping field that confines the molecules and the electrostatic field such that $\beta$ $\ne$ 0, $\pi$. In that case, the degeneracy of the $ \pm $M  states is lifted \textsuperscript{\cite{Bretislav1,Bretislav2}}. Alternatively, for molecules with a nuclear electric quadrupole moment, a superimposed magnetic field would lift the $ \pm $M degeneracy via the interaction between this moment and the magnetic moment generated by molecular rotation \textsuperscript{\cite{20-Bo Yan2013,31-S.Ospelkaus2010}}.

One potential application of polar molecules is in quantum computing, as originally proposed by DeMille two decades ago \textsuperscript{\cite{11-D. DeMille2002}}. Since then, many aspects and variants have been extensively studied, including for both diatomic linear molecules and symmetric top molecules \textsuperscript{\cite{12-Philippe Pellegrini2011,13-Jing Zhu2013,14-Zuo-Yuan Zhang2017,15-Zuo-Yuan Zhang2020,16-Wei12011,17-Wei22011,32-Micheli,33-Charron,34-kuz,35-ni,36-lics,37-YelinDeMille,38-Wei2010,39-Wei2016,40-kang-Kuen Ni2018}}. For linear molecules, the $\left| {0 0} \right\rangle$ and $\left| {1 0} \right\rangle$ pendular states are the most commonly used qubit states\textsuperscript{\cite{12-Philippe Pellegrini2011,13-Jing Zhu2013,15-Zuo-Yuan Zhang2020,16-Wei12011,32-Micheli,33-Charron,34-kuz,35-ni,36-lics,37-YelinDeMille,38-Wei2010,39-Wei2016,40-kang-Kuen Ni2018}}. For symmetric top molecules, different choices of qubit states have been explored\textsuperscript{\cite{17-Wei22011,14-Zuo-Yuan Zhang2017}}. In most cases, the Hamiltonian matrices are complex and no existing model can be used directly. In the meantime, spin systems are also considered to be a promising platform to implement a quantum computer. In fact, most of the work on quantum computers is based on spin systems (Note: superconducting loops are actually artificial spins)\textsuperscript{\cite{41-L.M.K.2000,42-J.Zhang2005,43-L.M.K2004,44-Kavita2000,45-Chiu2004,46-V.W.Scarola2005,47-F.B.M2009,48-M.Asoudeh2004,49-Meng2015,50-V.V.Aristov2004}} and Heisenberg model is the most popular model used in treating such systems\textsuperscript{\cite{45-Chiu2004,46-V.W.Scarola2005,47-F.B.M2009,48-M.Asoudeh2004,49-Meng2015}}. If we take our pseudo-spin states $\left|  \downarrow  \right\rangle$ and $\left|\uparrow\right\rangle$ as qubits states, then the two methods coincide. This opens up the prospect of directly transplanting methods and techniques developed for spins to polar molecules.

So far, most proposals for implementing quantum computing with polar molecules have been based on the gate model. Our new choices of qubits states also invite the possibility of adiabatic quantum computing  \textsuperscript{\cite{51-BRIW+14,52-SQVL14,53-VMBR+14,54-VMKO15,55-HJAR+15,56-KHZO+15,57-Ta,58-Gre,59-Ari,60-Sha,61-Ke}}. This follows primarily from the fact that the energy gap between the two qubit states $\Delta E = E_1-E_0$ can be arbitrarily tuned from 0 to 3.7B (see Figure  \ref{fig1}) by changing electric fields. For adiabatic quantum computing, the tunable energy gap between $|0\rangle$ and $|1\rangle$ need to be large compared with the interaction energy $V_{d-d}$. In this case, it is around $10^4$ times larger than the coupling energy $V_{d-d}$, which is far beyond the current limit that spin systems can achieve  \textsuperscript{\cite{57-KYNH+15}}. Moreover, one requirement for adiabatic quantum computing is that the energy gap between the ground and the first excited state be maintained during the adiabatic evolution such that no phase transitions could occur. This requirement is also satisfied, since for a practical coupling constant ($\Omega/B<10^{-2}$) during the adiabatic process of reducing the electric field, the entire polar molecular system remains in the ferromagnetic phase, without undergoing any phase transition (see Figure \ref{fig8}).

\section*{ACKNOWLEDGEMENTS}
We are grateful for support from National Natural Science Foundation of China (Grant Nos. 11974113 and 11674098). SK would like to acknowledge the support of the National Science Foundation under award number 1955907. BF gratefully acknowledges the hospitality of John Doyle and Hossein Sadeghpour during his stay at Harvard Physics and at the Harvard-Smithsonian Institute for Theoretical Atomic, Molecular, and Optical Physics (ITAMP).

\end{document}